\documentclass[a4paper,twoside,12pt]{article}
\input{obsjkt.sty}

\newcommand{\Msun}{\ensuremath{\,{\rm M}_\odot}}                  
\newcommand{\Rsun}{\ensuremath{\,{\rm R}_\odot}}                  
\newcommand{\rhosun}{\ensuremath{\,\rho_\odot}}                   
\newcommand{\Teff}{\ensuremath{T_{\rm eff}}}                      
\newcommand{\degr}{\ensuremath{^\circ}}                           
\renewcommand{\kms}{\,km\,s$^{-1}$}                               
\newcommand{\etal}{\textit{et al.}}                               

\newcommand{\kepler}{\textit{Kepler}}
\newcommand{\tess}{\textit{TESS}}
\newcommand{\gaia}{\textit{Gaia}}

\newcommand{\Msunnom}{\hbox{$\mathcal{M}^{\rm N}_\odot$}}
\newcommand{\Rsunnom}{\hbox{$\mathcal{R}^{\rm N}_\odot$}}
\newcommand{\Lsunnom}{\hbox{$\mathcal{L}^{\rm N}_\odot$}}

\begin{document} 

\OBSheader{Rediscussion of eclipsing binaries: AN Cam}{J.\ Southworth}{2021 Jun}

\OBStitle{Rediscussion of eclipsing binaries. Paper IV. \\ The evolved G-type system AN\,Camelopardalis}

\OBSauth{John Southworth}

\OBSinstone{Astrophysics Group, Keele University, Staffordshire, ST5 5BG, UK}

\OBSabstract{AN\,Cam is a little-studied eclipsing binary containing somewhat evolved components in an orbit with a period of 21.0\,d and an eccentricity of 0.47. A spectroscopic orbit based on photoelectric radial velocities was published in 1977. AN\,Cam has been observed using the \tess\ satellite in three sectors: the data were obtained in long-cadence mode and cover nine eclipses. By modelling these data and published radial velocities we obtain masses of $1.380 \pm 0.021$\Msun\ and $1.402 \pm 0.025$\Msun, and radii of $2.159 \pm 0.012$\Rsun\ and $2.646 \pm 0.014$\Rsun. We also derive a precise orbital ephemeris from these data and recent times of minimum light, but find that the older times of minimum light cannot be fitted assuming a constant orbital period. This could be caused by astrophysical or instrumental effects; forthcoming \tess\ observations will help the investigation of this issue. We use the \gaia\ EDR3 parallax and optical/infrared apparent magnitudes to measure effective temperatures of $6050 \pm 150$\,K and $5750 \pm 150$\,K: the primary star is hotter but smaller and less massive than its companion. A comparison with theoretical models indicates that the system has an approximately solar chemical composition and an age of 3.3\,Gyr. Despite the similarity of their masses the two stars are in different evolutionary states: the primary is near the end of its main-sequence lifetime and the secondary is now a subgiant. AN\,Cam is a promising candidate for constraining the strength of convective core overshooting in 1.4\Msun\ stars.}


\section*{Introduction}

Although the current generation of theoretical stellar models \cite{Bressan+12mn,Ekstrom+12aa,Paxton+11apjs,Dotter16apjs} provides a sophisticated and -- in many cases -- impressively accurate description of the behaviour of stars, there remain multiple effects which are still not properly understood. These include rotation in high-mass stars \cite{Venn+02apj,Talon++06apj,Maeder09book}, convective core overshooting \cite{ClaretTorres18apj,ConstantinoBaraffe18aa}, mixing length \cite{Graczyk+16aa}, opacities \cite{Imbriani+04aa,Pignatari+13apj,VillanteSerenelli21}, and the influence of magnetic activity in low-mass stars \cite{Lopez07apj,Kraus+11apj,Spada+13apj}. Additional constraints on these phenomena are needed via empirical determinations of the basic physical properties of stars. Those that have completed their main-sequence lifetime and are experiencing the faster evolutionary states that follow are most valuable because the predictions of theoretical models are more sensitive to the physics included in the models.

One of the primary sources of such empirical measurements is eclipsing binary stars \cite{Andersen91aarv,Torres++10aarv}, as their physical properties can be determined using only photometry, spectroscopy and geometry. Detached eclipsing binaries (dEBs) are the most valuable because the two components can be assumed to have evolved as single stars, and thus allow the measurements of the properties of two stars of different mass but the same age and initial chemical composition. dEBs have been used, among other things, to calibrate empirical mass--radius--temperature--age relations \cite{Torres++10aarv,Enoch+10aa,Me11mn,Moya+18apjs}, investigate the treatment of mixing length and core overshooting in theoretical models \cite{Andersen++90apj,Pols+97mn,ClaretTorres16aa,Tkachenko+20aa}, and study stellar chemical evolution \cite{PavlovskiMe09mn,Pavlovski+09mn,Pavlovski++18mn}.


\begin{table}[t]
\caption{\em Basic information on AN\,Cam \label{tab:info}}
\centering
\begin{tabular}{lll}
{\em Property}                      & {\em Value}           & {\em Reference}                   \\[3pt]
Henry Draper designation            & HD 24906              & \cite{CannonPickering18anhar2}    \\
\textit{Tycho} designation          & TYC 4514-8-1          & \cite{Hog+00aa}                   \\
\textit{Gaia} EDR3 designation      & 551506893532348544    & \cite{Gaia20aa}                   \\
\textit{Gaia} parallax              & $3.208 \pm 0.015$ mas & \cite{Gaia20aa}                   \\
$B$ magnitude                       & $10.30 \pm 0.024$     & \cite{Hog+00aa}                   \\
$V$ magnitude                       & $ 9.69 \pm 0.03$      & \cite{Hog+00aa}                   \\
$J$ magnitude                       & $8.396 \pm 0.024$     & \cite{Cutri+03book}               \\
$H$ magnitude                       & $8.098 \pm 0.021$     & \cite{Cutri+03book}               \\
$K_s$ magnitude                     & $8.041 \pm 0.019$     & \cite{Cutri+03book}               \\
Spectral type                       & F8                    & \cite{CannonPickering18anhar2}    \\[10pt]
\end{tabular}
\end{table}

\section*{AN\,Camelopardalis}

In this work we present an analysis of AN\,Cam (Table\,\ref{tab:info}), a dEB containing two stars with masses near 1.4\Msun\ but with significantly different radii and effective temperature (\Teff) values. AN\,Cam has an eccentric orbit with a relatively long period of 21.00\,d, meaning that tidal effects are weak and thus the two stars have evolved in isolation since their formation. The current analysis is part of our efforts to extend the number of objects in DEBCat\footnote{\texttt{https://www.astro.keele.ac.uk/jkt/debcat/}} (the Detached Eclipsing Binary Catalogue), a compilation of dEBs with masses and radii measured to precisions of 2\% or better \cite{Me15debcat}; see Southworth\cite{Me20obs,Me21obs1,Me21obs2} for further discussion.

Very few studies of AN\,Cam have been published. Its eclipsing nature was discovered by Strohmeier \& Knigge \cite{StrohmeierKnigge60vebam2} and times of mid-eclipse have been obtained by several authors since \cite{Brelstaff82baa,Samolyk10javso,Hubscher11bavsm,Hubscher++12ibvs,HubscherLehmann15ibvs,Kim+18apjs}. The only other publication of note is that of Imbert\cite{Imbert87aas}, who presented a double-lined spectroscopic orbit for the system based on a total of 73 radial velocity (RV) measurements. These were obtained using the \textit{CORAVEL} spectrometer at l'Observatoire de Haute-Provence \cite{Baranne++79va}, in which a physical mask was used to directly observe the cross-correlation functions and thus the RVs of the stars \cite{Griffin67apj}. These RVs will be used below as they are crucial to the measurement of the masses and radii of the stars in the AN\,Cam system.


\section*{Observational material}

\begin{figure}[t] \centering \includegraphics[width=\textwidth]{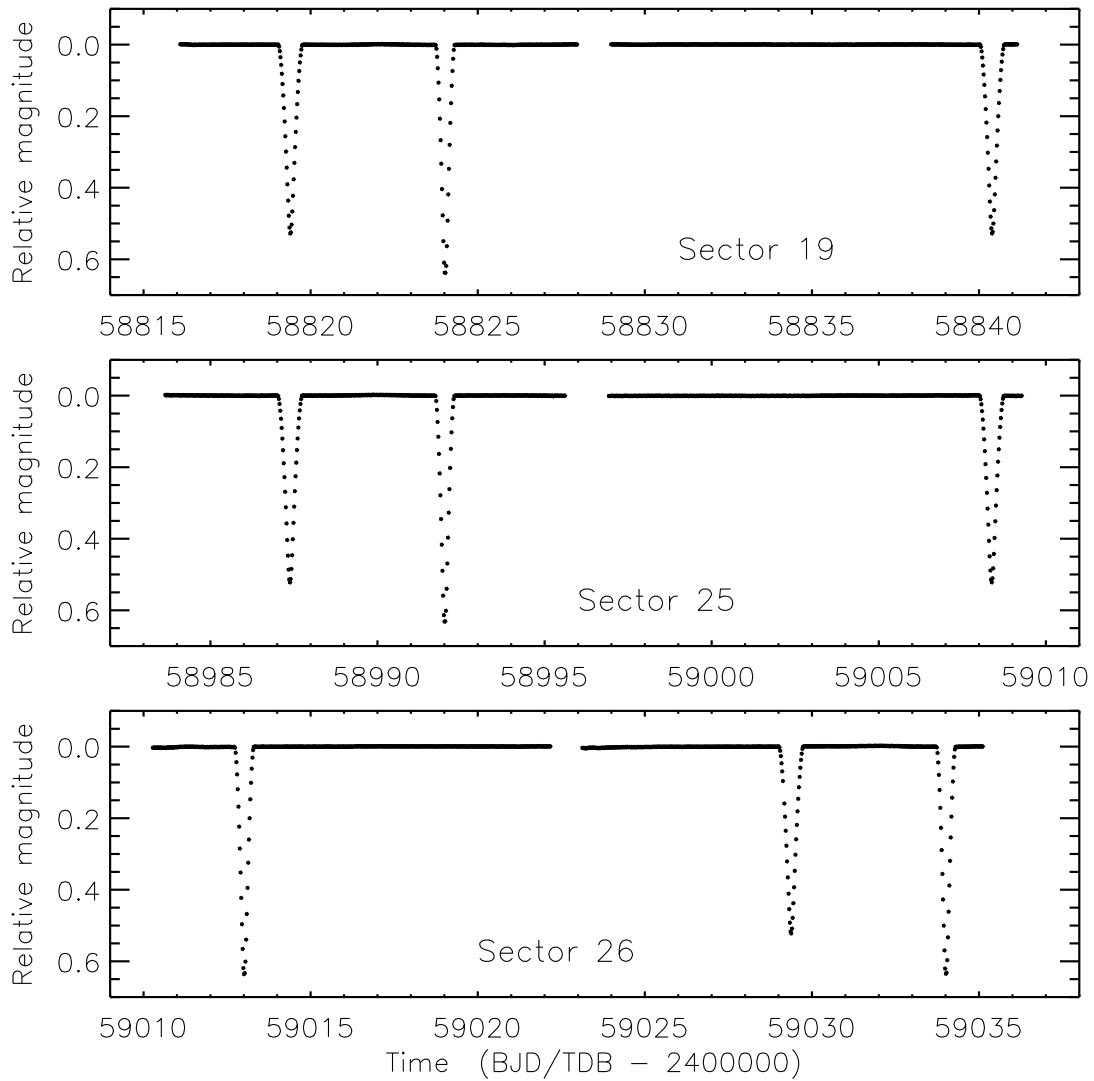} \\
\caption{\label{fig:time} \tess\ full-frame-image photometry of AN\,Cam.
The three plots show the data from sectors 19, 25 and 26, respectively.}
\end{figure}

The light curve studied in this work comes from the NASA \tess\ satellite \cite{Ricker+15jatis}. AN\,Cam was observed three times: in Sector 19 (2019/11/27 to 2019/12/24), Sector 25 (2020/05/13 to 2020/06/08) and Sector 26 (2020/06/08 to 2020/11/04). In contrast to previous papers of this series, the target was not selected for short-cadence observations. Its light curve was therefore extracted from the full-frame images using the \textsc{Lightkurve} package \cite{Lightkurve18}. AN\,Cam is also planned to be observed in Sectors 52 and 53 (2022 May--July) so future observations of a similar quality will become available (assuming \tess\ remains operational).

%
%

The end result of this process was a light curve containing 3463 datapoints with a sampling rate of 1800\,s (see Fig.\,\ref{fig:time}). As the dEB is well-detached and shows negligible proximity effects, the data outside eclipse contain no useful information. We therefore trimmed all observations more than 1.5 eclipse durations from the midpoint of an eclipse, thus retaining 803 datapoints for subsequent analysis.



\section*{Light curve analysis}

\begin{figure}[t] \centering \includegraphics[width=\textwidth]{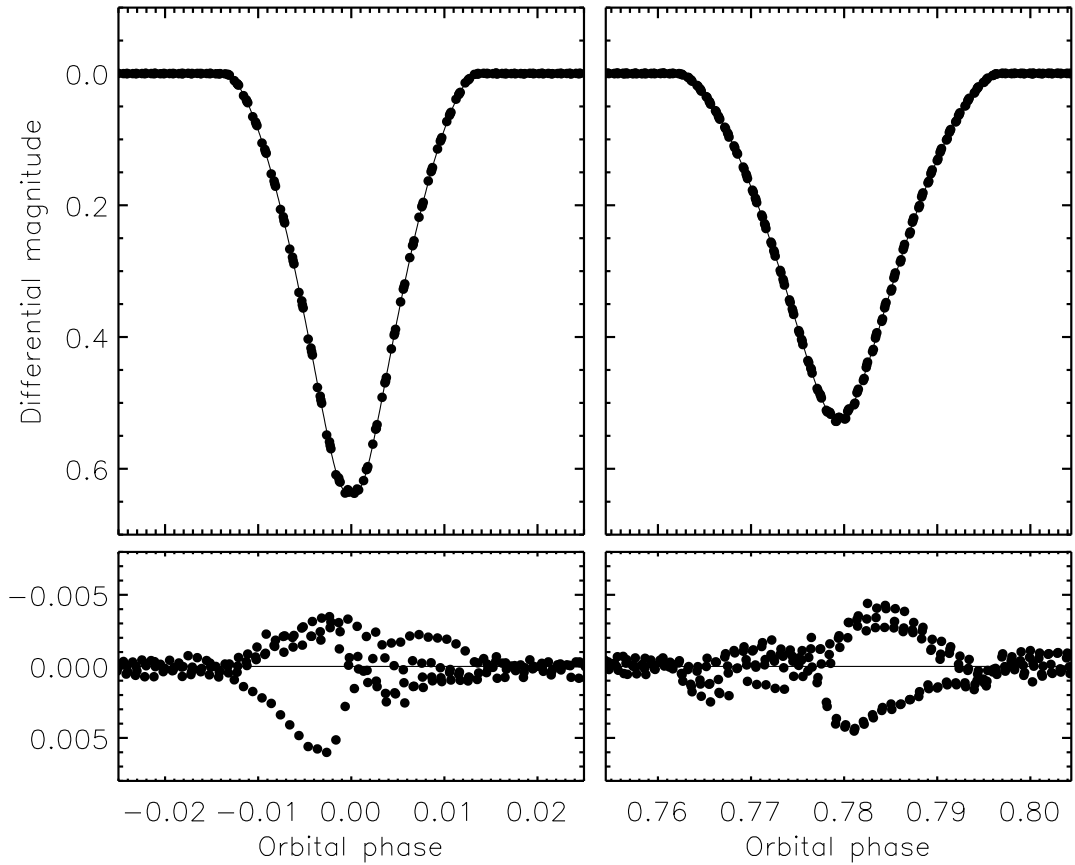} \\
\caption{\label{fig:tess} The \tess\ light curve of AN\,Cam (filled circles) around the primary
(left) and secondary (right) eclipses. The {\sc jktebop} best fit is shown using a solid
line. The lower panels show the residuals of the fit on a larger scale.} \end{figure}

We adopt the standard definition that the primary star is the one eclipsed at primary eclipse (at phase 0), and primary eclipse is deeper than secondary eclipse. It will be shown below that in the case of AN\,Cam this means the primary star is hotter than the secondary, but is smaller, less massive, and less bright. We refer to the primary as star A and the secondary as star B.

AN\,Cam has a relatively long orbital period so the stars are approximately spherical. We therefore elected to model the \tess\ data using version 41 of the {\sc jktebop}\footnote{\texttt{http://www.astro.keele.ac.uk/jkt/codes/jktebop.html}} code \cite{Me++04mn2,Me13aa}. {\sc jktebop} is fast and flexible, and has been found to be in good agreement with other codes for well-detached EBs \cite{Maxted+20mn}.

The radii of the stars are parameterised in {\sc jktebop} as the sum and ratio of the fractional radii ($r_{\rm A}+r_{\rm B}$ and $k = \frac{r_{\rm B}}{r_{\rm A}}$, where $r_{\rm A} = \frac{R_{\rm A}}{a}$ and $r_{\rm B} = \frac{R_{\rm B}}{a}$ are the fractional radii, $R_{\rm A}$ and $R_{\rm B}$ are the true radii of the stars, and $a$ is the semimajor axis of the relative orbit) and both quantities were fitted. AN\,Cam has an eccentric orbit, as can be seen from the fact that the secondary eclipse is not at phase 0.5 and is not the same duration as primary eclipse. This was accounted for by fitting for the quantities $e\cos\omega$ and $e\sin\omega$ where $e$ is the orbital eccentricity and $\omega$ is the argument of periastron. Other fitted parameters were the orbital inclination $i$, the central surface brightness ratio $J$, the orbital period $P$, and a reference time of primary mid-eclipse $T_0$. Attempts to fit for third light yielded negative values for this quantity, so we fixed its value at zero.

Limb darkening was included in the model using the quadratic law \cite{Kopal50} with coefficients for the \tess\ passband from Claret\cite{Claret18aa}. The two stars have similar surface gravities and \Teff\ values so we assumed the same coefficients for each star. We fitted for the linear coefficient but fixed the quadratic coefficient; the two coefficients are strongly correlated \cite{Me++07aa,Me08mn} so this does not introduce a significant dependence on theoretical models of stellar atmospheres.

\begin{figure}[t] \centering \includegraphics[width=\textwidth]{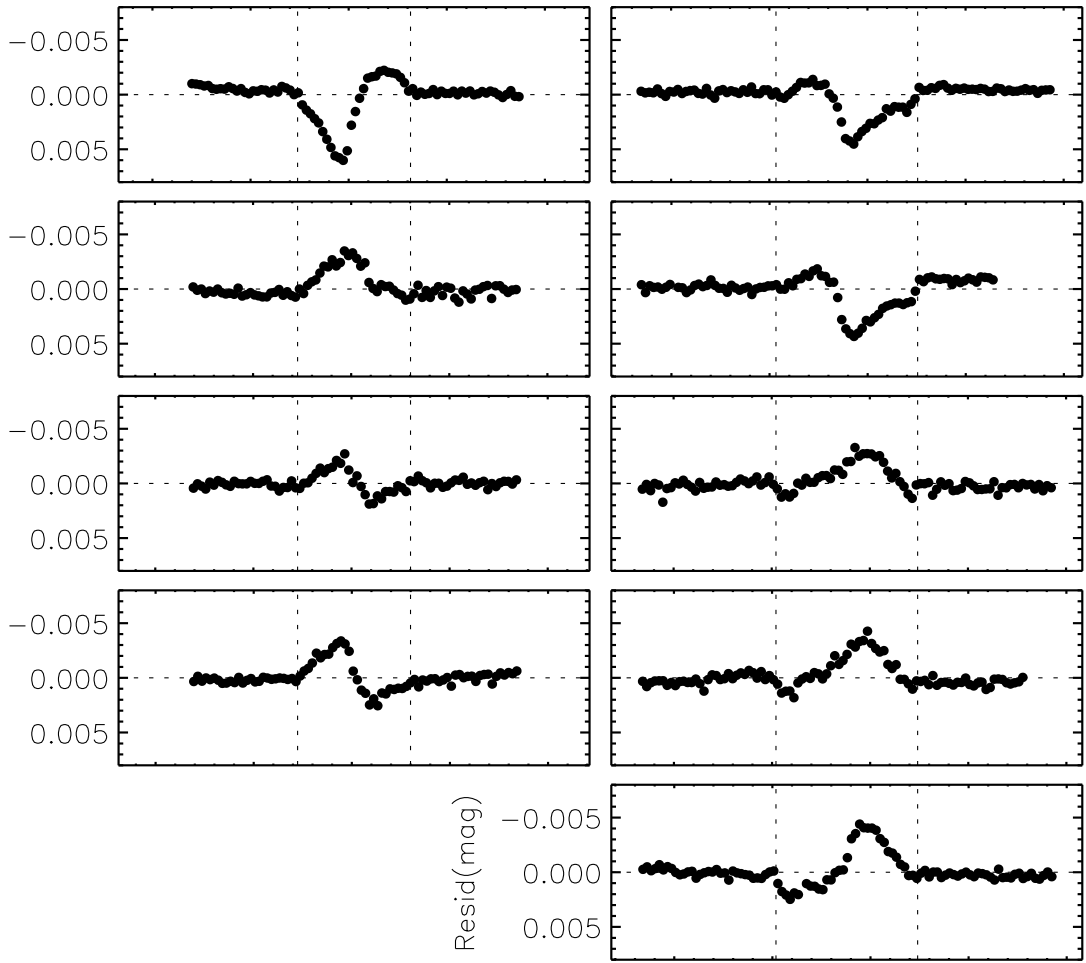} \\
\caption{\label{fig:resid} Residuals of the best fit to the \tess\ data around each
eclipse (filled circles). The primary eclipses are shown in the left panels and
the secondary eclipses in the right panels. Each panel covers a time interval of
2.4\,d centred on a time of mid-eclipse. The horizontal dashed lines indicate
a residual of zero and the vertical dashed lines show the start and end points
of the eclipses.} \end{figure}

A first fit to the \tess\ light curve was obtained by including a first-order polynomial for each eclipse to normalise it to zero differential magnitude. The relatively long duration of individual datapoints could lead to smearing of the eclipse shapes even at this orbital period, so the theoretical light curve was numerically integrated by averaging the model for each datapoint \cite{Me11mn}. Each average was calculated from five samples evenly covering 1800\,s. The secondary eclipse occurs at phase 0.7794.

\begin{figure}[t] \centering \includegraphics[width=\textwidth]{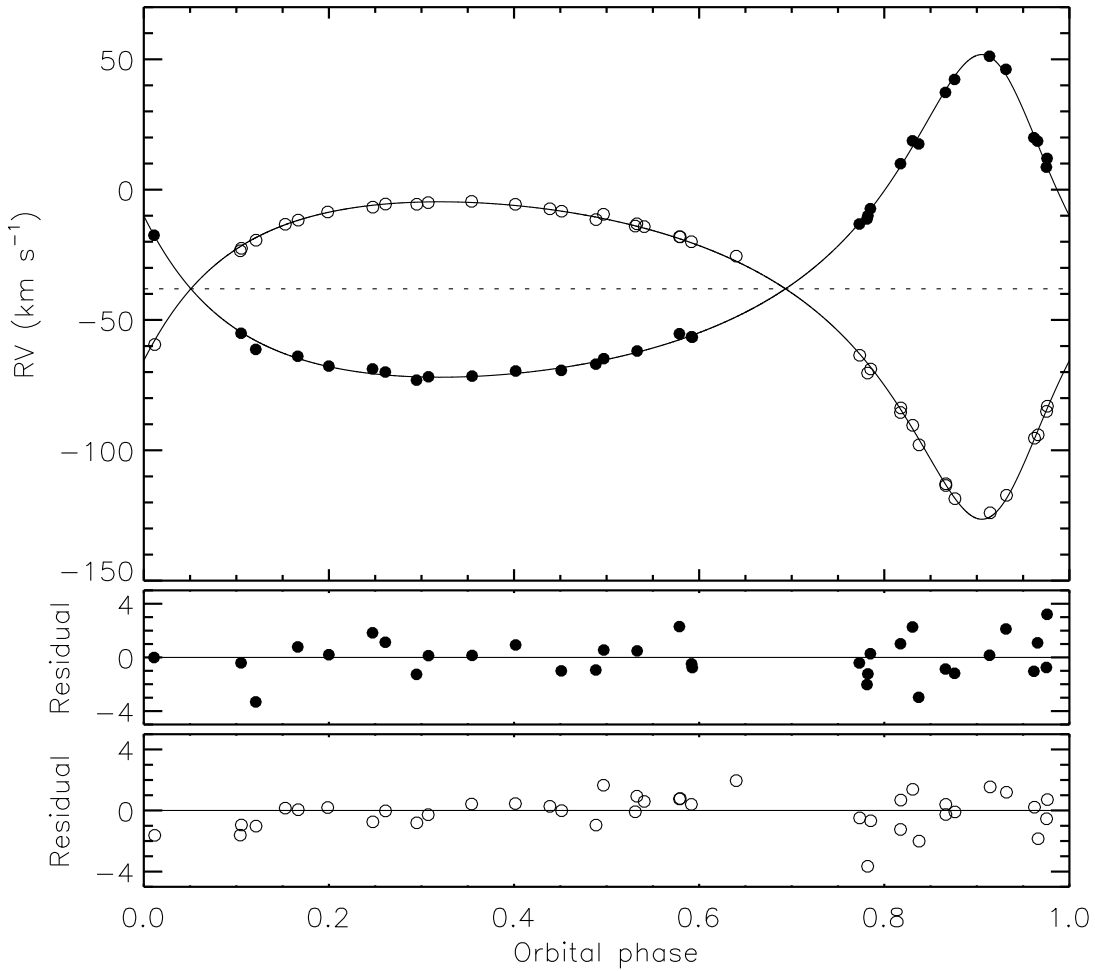} \\
\caption{\label{fig:rv} Spectroscopic orbit of AN\,Cam using the RVs from
Imbert\cite{Imbert87aas}. RVs of the primary and secondary stars are shown
with filled and open circles, respectively. The best fits from {\sc jktebop}
are shown using solid lines and the systemic velocity is indicated with a
dotted line. The lower panels show the residuals of the fit.} \end{figure}

The best fit was found to be a good match to the \tess\ data, but with significant deviations during eclipse as large as 0.5\,mmag (see Fig.\,\ref{fig:tess}). These could plausibly be explained by the presence of a third light that changes over the course of the observations -- possible given that the \tess\ data of AN\,Cam are from two different cameras -- or by surface inhomogeneities (starspots) that evolve over time. Closer inspection of the residuals of the fit (see Fig.\,\ref{fig:resid}) shows that the residuals change with the orbit and are not just miniature versions of the eclipses, so are consistent with spot activity but not with an erroneous third light value. The residuals are seen in both primary and secondary eclipse, which means that spots are present on both stars. No evidence for periodicity due to spot rotation is found in the TESS observations, implying either a slow rotation or a complex spot distribution on the components' surfaces. A colleague suggested the possibility of apsidal motion, but this effect is expected to be too weak to affect our analysis because of the relatively long orbital period, small fractional radii (even at periastron) and the fact that the stars are moderately evolved so are quite centrally condensed.

\begin{table} \centering
\caption{\em Best {\sc jktebop} fit to the \tess\ light curve of AN\,Cam. The RVs from
Imbert\cite{Imbert87aas} and four times of minimum light (Table\,\ref{tab:tmin}) were
included in the fit. The uncertainties are 1$\sigma$. The same limb darkening
coefficients were used for both stars. The uncertainties in the systemic velocities
do not include any transformation onto a standard system. \label{tab:lc}}
\begin{tabular}{lr@{\,$\pm$\,}l}
{\em Parameter}                           & \multicolumn{2}{c}{\em Value}    \\[3pt]
{\it Fitted parameters:} \\
Primary eclipse time (BJD/TDB)            & 2458992.01472   & 0.00095         \\
Orbital period (d)                        &      20.998420  & 0.000012        \\
Orbital inclination (\degr)               &      89.213     & 0.013           \\
Sum of the fractional radii               &       0.10665   & 0.00010         \\
Ratio of the radii                        &       1.2256    & 0.0015          \\
Central surface brightness ratio          &       0.8368    & 0.0008          \\
Linear limb darkening coefficient         &       0.190     & 0.011           \\
Quadratic limb darkening coefficient      & \multicolumn{2}{c}{~~~~~~~~~~0.22 (fixed)}  \\
$e\cos\omega$                             &       0.45176   & 0.00009         \\
$e\sin\omega$                             &       0.1208    & 0.0009          \\
Velocity amplitude of star A (\kms)       &      61.89      & 0.52            \\
Velocity amplitude of star B (\kms)       &      60.89      & 0.38            \\
Systemic velocity of star A (\kms)        &   $-$38.04      & 0.28            \\
Systemic velocity of star B (\kms)        &   $-$38.05      & 0.16            \\[3pt]
{\it Derived parameters:} \\
Fractional radius of star A               &       0.047920  & 0.000077        \\
Fractional radius of star B               &       0.058739  & 0.000032        \\
Orbital eccentricity                      &       0.4676    & 0.0002          \\
Argument of periastron (\degr)            &      14.97      & 0.11            \\
Light ratio                               &       1.2570    & 0.0041          \\
\end{tabular}
\end{table}

Once the \tess\ data were adequately modelled, we added in the RVs from Imbert\cite{Imbert87aas}. These comprise 33 measurements for star A and 40 for the brighter star B. The velocity amplitudes and systemic velocities of the two stars were included as fitted parameters. The fitted spectroscopic orbit is shown in Fig.\,\ref{fig:rv}. We also included some published times of minimum light to further constrain the orbital ephemeris of the system; these will be discussed below. The final results are shown in Table\,\ref{tab:lc}.


\section*{Error analysis}

The systematic trends in the residuals visible in Fig.\,\ref{fig:resid} are concerning from the viewpoint of error analysis, as they break the standard assumption that the datapoints are \emph{iid} (independent and identically distributed). The uncertainties in the fitted and derived parameters were therefore obtained in three different ways, and the largest of the three options was retained for each parameter.

The first method used for estimating the uncertainties was the Monte Carlo algorithm implemented in \textsc{jktebop} \cite{Me++04mn}, which assumes \emph{iid} datapoints but does account for correlations between parameters. The second method was the residual-permutation algorithm in \textsc{jktebop} \cite{Me08mn}, which cyclically permutes the residuals through the data and then refits, so can capture the effects of correlated noise on the determinacy of the solution. The third method was to perform a fit to each of the three sectors of \tess\ data separately and deduce uncertainties from the scatter of the three parameter values that resulted.

In the case of AN\,Cam we found that the Monte-Carlo errorbars were the largest for most of the fitted and derived parameters. The residual-permutation errorbars were largest for the surface brightness ratio and the velocity amplitudes. The separate-fit errorbars were greatest for the ratio of the radii, the fractional radius of star A, and the light ratio of the stars. The parameters reported in Table\,\ref{tab:lc} are from the joint fit to the full data, and the errorbars are the largest of the three options for each parameter. The future observations from \tess\ should allow both the precision and accuracy of the measured parameters to be improved.


\section*{Orbital ephemeris}

The \tess\ data provide a good measurement of the orbital period of AN\,Cam, because they cover a time interval of 216\,d and the eclipses are sharp and deep. However, the RVs were obtained 40 years prior so we sought additional constraints on the orbital ephemeris of the system. We found a total of nine times of mid-eclipse from seven sources in the literature, of which three times had an associated uncertainty.

On adding these into the {\sc jktebop} solution we found that the fit to the \tess\ data was severely compromised, and that most of the eclipse times deviated from the best-fitting linear ephemeris by many times their uncertainties. After extensive but inconclusive investigations we resolved to include only the four most recent eclipse times and to apply a uniform uncertainty of 0.01\,d to them. Our justifications for this approach are: (1) that the earlier eclipse times differ by at least 0.02\,d and as much as 0.26\,d, and it is best to reject them \textit{en masse} than to pick those which happen to provide a better agreement\footnote{It is worth remembering Merrill's theorem: \emph{once discrepant measurements are rejected the remainder will be found to agree well.}}; and (2) the extremely small quoted uncertainty in one of the remaining eclipse times is hard to justify for this system and is not supported by such a level of agreement with the fitted ephemeris.

This is clearly an unsatisfactory situation, and the prospective future observations from \tess\ in Sectors 52 and 53 will help improve it. We rest our own analysis primarily on the \tess\ data, which are incomparably better than any previous light curves of AN\,Cam, and on the good agrement found with the RVs from Imbert\cite{Imbert87aas}. Although the large residuals in the earlier data hint at orbital period changes or apsidal motion in this system, it is more plausible that they are compromised by the difficulty of measuring precise timings in eclipses much longer than a typical observing night (13.8\,hr for the primary eclipse and 17.3\,hr for the secondary), deformation of the eclipse shapes by spot activity, and/or clock errors in the equipment used.

Inclusion of the four most recent eclipse times significantly improves the measurement of the orbital ephemeris for AN\,Cam, and has a negligible effect on the other parameters of the {\sc jktebop} fit. The eclipse times and fits are given in Table\,\ref{tab:tmin}.

\begin{table} \centering
\caption{\em Times of published mid-eclipse for AN\,Cam and their residuals versus the best fit reported
in the current work. Asterisks indicate times not included in the final best fit. The orbital cycle is
an integer for primary eclipses and a half-integer for secondary eclipses. \label{tab:tmin}}
\begin{tabular}{llllll}
{\em Orbital} & {\em Eclipse} & {\em Published } & {\em Adopted } & {\em Fitted} & {\em Source} \\
{\em cycle} & {\em time (BJD)} & {\em uncertainty (d)} & {\em uncertainty (d)} & {\em time (BJD)} & \\[3pt]
$-$1571   & 2426003.47   &        &      & 26003.4969 & \cite{StrohmeierKnigge60vebam2} \\
 $-$665.5 & 2445023.32   &        &      & 45023.4328 & \cite{Brelstaff82baa}           \\
 $-$634.5 & 2445674.23   &        &      & 45674.3839 & \cite{Brelstaff82baa}           \\
 $-$355.5 & 2451532.683  &        &      & 51532.9430 & \cite{Kim+18apjs}               \\
 $-$355   & 2451537.605  &        &      & 51537.5756 & \cite{Kim+18apjs}               \\
 $-$192.5 & 2454955.6932 &        & 0.01 & 54955.6855 & \cite{Samolyk10javso}           \\
 $-$162   & 2455590.2891 & 0.0079 & 0.01 & 55590.2707 & \cite{Hubscher11bavsm}          \\
 $-$160.5 & 2455627.6431 & 0.0065 & 0.01 & 55627.6349 & \cite{Hubscher++12ibvs}         \\
 $-$100.5 & 2456887.5481 & 0.0001 & 0.01 & 56887.5401 & \cite{HubscherLehmann15ibvs}    \\
\end{tabular}
\end{table}


\section*{Physical properties of AN\,Cam}

Although {\sc jktebop} provided the masses and radii of the components of AN\,Cam measured from the \tess\ light curve and Imbert\cite{Imbert87aas} RVs, we used the {\sc jktabsdim} code \cite{Me++05aa} to calculate the physical properties of the system in order to include quantities such as \Teff, luminosity and distance. The results were in good agreement with those from {\sc jktebop}, but the uncertainties were greater in some cases due to the adoption of the largest of three alternative error estimates for each measured parameter (see above). The precision to which the radii are measured is limited by the RV observations, which set the scale of the system, and not by the \tess\ light curve.

\begin{table} \centering
\caption{\em Physical properties of AN\,Cam. Units superscripted with an `N'
are defined by IAU 2015 Resolution B3 \cite{Prsa+16aj}. \label{tab:absdim}}
\begin{tabular}{lr@{\,$\pm$\,}lr@{\,$\pm$\,}l}
{\em Parameter}        & \multicolumn{2}{c}{\em Star A} & \multicolumn{2}{c}{\em Star B} \\[3pt]
Mass ratio                                  & \multicolumn{4}{c}{$1.016 \pm 0.011$}      \\
Semimajor axis of relative orbit (\Rsunnom) & \multicolumn{4}{c}{$45.05 \pm 0.24$}       \\
Mass (\Msunnom)                             &   1.380 & 0.021       &   1.402 & 0.025    \\
Radius (\Rsunnom)                           &   2.159 & 0.012       &   2.646 & 0.014    \\
Surface gravity ($\log$[cgs])               &  3.9095 & 0.0030      &  3.7400 & 0.0037   \\
Density ($\!$\rhosun)                       &  0.1371 & 0.0010      &  0.0757 & 0.0004   \\
Synchronous rotational velocity (\kms)      &   5.201 & 0.029       &   6.376 & 0.034    \\
Effective temperature (K)                   &    6050 & 150         &    5750 & 150      \\
Luminosity $\log(L/\Lsunnom)$               &   0.750 & 0.043       &   0.839 & 0.046    \\
$M_{\rm bol}$ (mag)                         &    2.86 & 0.11        &    2.64 & 0.11     \\
~ \\
~ \\
~ \\
\end{tabular}
\end{table}

The two important quantities unavailable from the preceding analysis are the \Teff\ values of the stars. These could be constrained from the distance ($311.7 \pm 1.4$\,pc based on the parallax from \gaia\ EDR3\cite{Gaia20aa}), the $BV$ and $JHK$ apparent magnitudes (Table\,\ref{tab:info}), and the light ratio in the \tess\ passband (Table\,\ref{tab:lc}) of the system. We first obtained a value of $E_{B-V} = 0.06 \pm 0.04$ for the interstellar reddening of the system\footnote{\texttt{https://stilism.obspm.fr}} from Lallement \etal\ \cite{Lallement+14aa,Lallement+18aa}. We then determined the ratio of the \Teff\ values of the stars using theoretical spectra from the {\sc atlas9} model atmospheres \cite{Castelli++97aa}, the light ratio in the \tess\ passband, and the \tess\ passband response function \cite{Ricker+15jatis}. The \Teff\ values of the stars were then iteratively determined using {\sc jktabsdim} and the bolometric corrections from Girardi \etal\ \cite{Girardi+02aa} in order to match the distance to the system known from the \gaia\ parallax. We determined conservative uncertainties based on the maximum possible perturbation that could be applied to the \Teff\ values and still match the \gaia\ parallax whilst retaining a consistent distance between the different optical--infrared passbands for which apparent magnitudes are available. The final results are given in Table\,\ref{tab:absdim}. The \Teff\ values imply a spectral type of F9\,V + G2\,IV for the system, using the calibration of Pecaut \& Mamajek \cite{PecautMamajek13apjs}. The atmospheric properties of the system could be measured better by a detailed analysis of high-resolution spectra of the dEB.


\section*{The evolutionary status of AN\,Cam}

\begin{figure}[t] \centering \includegraphics[width=\textwidth]{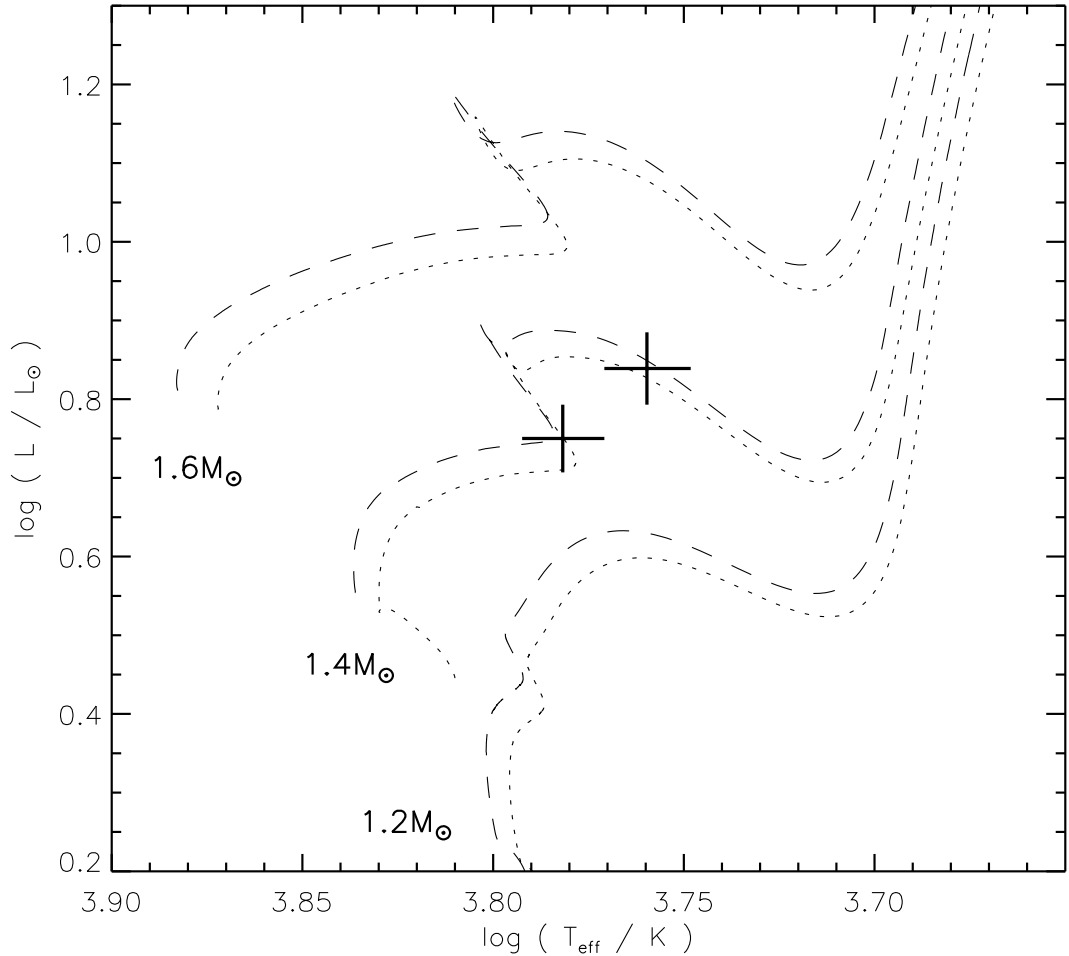} \\
\caption{\label{fig:hrd} Hertzsprung-Russell diagram showing the components of AN\,Cam
(solid crosses) and selected predictions from the PARSEC models \cite{Bressan+12mn}
(broken lines) beginning at the zero-age main sequence. Models for 1.2, 1.4 and 1.6\Msun\
are shown (labelled). In each case the dashed line is for a metal abundance of $Z=0.017$
and the dotted line is for $Z=0.020$.} \end{figure}

We have made a first preliminary comparison between the physical properties of the AN\,Cam system and the predictions of theoretical stellar models. A detailed analysis should be performed once precise spectroscopic \Teff\ and chemical abundance measurements are available. We chose to use the PARSEC models \cite{Bressan+12mn} for the current analysis, and restricted our comparison to chemical compositions near solar.

Fig.\,\ref{fig:hrd} shows a Hertzsprung-Russell diagram with the components of AN\,Cam and the predictions for a subset of the available PARSEC models. It can be seen that the evolutionary tracks for 1.4\Msun\ agree with the measured \Teff\ values and luminosities of the stars, and for both of the metal abundances shown ($Z=0.017$ and $Z=0.020$). To infer the age of the system we plotted the observed and theoretical data in mass--radius and mass--\Teff\ diagrams (not shown). A good agreement was found for an age of 3210\,Myr (for $Z=0.017$) or 3380\,Myr ($Z=0.020$). The formal uncertainty in these ages is only $\pm$10\,Myr, which is dwarfed by the systematic errors in the theoretical models (see Paper\,I \cite{Me20obs}).

Fig.\,\ref{fig:hrd} shows that the two components of the dEB are in different evolutionary states, with star A being near the terminal-age main sequence and star B having already passed this to become a subgiant. The system is comparable to the AI\,Phe \cite{Andersen+88aa,KirkbyKent+16aa,Maxted+20mn} and V501\,Her \cite{LacyFekel14aj} systems, but with the advantage that the masses of the stars are more similar despite the different evolutionary states\footnote{The masses of the two components differ by 1.6\% for AN\,Cam, 4.1\% for AI\,Phe and 4.6\% for V501\,Her.}. These systems, together with others such as RT\,CrB \cite{SabbyLacy03aj} and CF\,Tau \cite{Lacy++12aj} where both components have left the main sequence, are important tests of the treatment of convective mixing in theoretical stellar models \cite{Torres+14aj}.


\section*{Summary}

AN\,Cam has been known to be an eclipsing binary for over 60\,yr \cite{StrohmeierKnigge60vebam2} and a high-quality spectroscopic orbit exists \cite{Imbert87aas}, but it previously lacked a detailed photometric analysis. The light curve obtained by the \tess\ satellite has allowed this gap to be filled. We modelled the \tess\ data simultaneously with the RVs from Imbert \cite{Imbert87aas} and determined the masses to precisions of 1.5\% and 1.8\%, for star A and star B respectively, and the radii to precisions of 0.5\% for both stars. Published times of minimum light are mutually contradictory, suggesting the possibility of orbital period variations in the system or instrumental difficulties with the measurement of the times of midpoint of such long eclipses. The \Teff\ values of the stars were deduced from their light ratio in the \tess\ passband, the \gaia\ EDR3 parallax and apparent magnitudes of the system.  Both components show evidence for spot activity.

AN\,Can was found to contain two stars of very similar mass (1.38\Msun\ and 1.40\Msun) but nevertheless quite different radii (2.16\Rsun\ and 2.65\Rsun), bound in a 21.0-d period orbit with significant eccentricity ($e = 0.47$). These properties are consistent with an age of 3.3\,Gyr, an approximately solar chemical composition, and differing evolutionary states for the two stars (main sequence for star A and subgiant for star B). These properties mean AN\,Cam has the potential to allow a discriminating test of stellar evolutionary models.

AN\,Cam will be observed by \tess\ again, in mid-2022. A detailed spectroscopic analysis would be valuable for improving the precision of the mass measurements, and for determining precise \Teff\ values and photospheric chemical abundances. A combined analysis of this and other subgiants in dEBs may then allow strong constraints to be placed on the strength of convective core overshooting implemented in the current generation of theoretical stellar evolutionary models.


\section*{Acknowledgements}

We are grateful to Zac Jennings for extracting the \tess\ light curve of AN\,Cam, and to Dariusz Graczyk for helpful comments.
This paper includes data collected by the \tess\ mission. Funding for the \tess\ mission is provided by the NASA's Science Mission Directorate.
This research made use of \textsc{Light-kurve}, a Python package for \kepler\ and \tess\ data analysis \cite{Lightkurve18}.
This work has made use of data from the European Space Agency (ESA) mission \gaia, processed by the \gaia\ Data Processing and Analysis Consortium (DPA). Funding for the DPAC has been provided by national institutions, in particular the institutions participating in the \gaia\ Multilateral Agreement.
The following resources were used in the course of this work: the NASA Astrophysics Data System; the SIMBAD database operated at CDS, Strasbourg, France; and the ar$\chi$iv scientific paper preprint service operated by Cornell University.


\bibliographystyle{obsmaga}

\end{document}